\documentclass[11pt,twoside,a4paper]{article}

\usepackage{latexsym}
\usepackage[dvips]{graphicx} 
\begin{document} 

\begin{center}
{\LARGE The $\nu=1$ Quantum Hall edge with a \\ realistic potential}
\\
\ \\
{\large Joachim Sj\"ostrand, Anders Eklund, Anders Karlhede}
\\
\textit{Department of Physics, Stockholm University,}\\
\textit{Stockholm Center for Physics, Astronomy and Biotechnology}\\
\textit{S-10691 Stockholm, Sweden}\\
(March 20, 2002)\\
\end{center}


\begin{abstract}
\noindent
We study the softening of the edge modes of the compact ferromagnetic $\nu=1$ Quantum Hall edge as the strength of the confining potential decreases. To obtain a realistic potential we use a displaced background charge plane. We find a phase transition into a spin textured edge for $\tilde{g}=\frac{g\mu_B B}{e^2 / \epsilon l} \leq 0.012$.\end{abstract}

\setcounter{page}{1}

\section{Introduction}

Since the discovery of the Quantum Hall (QH) effect, work by Halperin \cite{Hal} and, in particular, Wen \cite{Wen} has lead us to understand that there is novel low energy physics at the edge of an incompressible QH system. The predicted Luttinger liquid correlations at the edges have been studied extensively, leading to predictions that have considerable support in experiments \cite{Chang}.

In this paper we investigate the edge of the compact ferromagnetic $\nu=1$ QH system, when the potential confining the electrons to the sample models a realistic potential. This edge has two types of collective excitations bound to it: the much studied edge magnetoplasmon (EMP) \cite{Wen}, and the edge spin wave (ESW) \cite{KL5}. When the confining potential softens (diminishes in strength), both these excitations soften and eventually their energies become negative, indicating that the edge reconstructs by spin texture (TEX) or charge density wave (CDW) formation \cite{KL1,Oaknin,Brey,Oak2}. The spin textured edge has the same spin structure as a QH skyrmion \cite{Rezayi}.

An extensively used model for the confining potential is the $w$-model \cite{Chamon}, where the potential is obtained from a linearly decreasing, in-plane positive background charge distribution. Unfortunately, however, the $w$-model is non-generic, and can not be used to determine, e.g., where, in parameter space, the phase transition into the TEX edge is. A natural attempt to obtain a more realistic description is to use the potential from a charge distribution displaced perpendicularly from the electron gas \cite{AE}. However, this gives a potential that converges to its bulk value very slowly, which leads to numerical problems. Here we handle this by modifying the potential slightly.

We determine where the energies of the EMP and ESW modes become negative as functions of the strength of the confining potential and of $\tilde{g}=\frac{g\mu_B B}{e^2 / \epsilon l}$ - the ratio between the Zeeman and Coulomb energies. This gives a phase diagram for the $\nu=1$ QH edge. In particular, we find that as the potential softens the transition from the compact ferromagnetic $\nu=1$ edge is into a TEX edge for $\tilde{g} \leq \tilde{g}_c = 0.012$.

\section{The system and the excitations}

\begin{figure}[!htb]
\begin{center}
\includegraphics{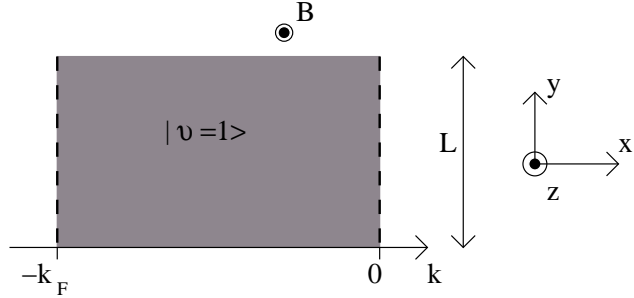}
\caption{\footnotesize{Geometry of our QH system.}}
\label{fig_1}
\end{center}
\end{figure}

The QH system that we consider is a two dimensional electron gas (2DEG) with a large magnetic field \textbf{B} applied perpendicular to the plane of the electrons. We restrict the electrons to the lowest Landau level (LLL). For the vector potential we use Landau gauge; $\textbf{A}=Bx \hat{\textbf{y}}$. The system has the geometry of a bar, with periodic boundary conditions in the $y$-direction. The circumference of this effective cylinder is $L$, and the number of electrons in the gas is $N$. The electron wave functions in the LLL are (with the magnetic length $l=\sqrt{\hbar c/eB}=1$)
\begin{equation}
\label{psi}
\psi_{k}(x,y) = \frac{1}{\sqrt{\pi^{1/2} L}} e^{iky}e^{-\frac{1}{2}(x-k)^2}.
\end{equation}
The quantum numbers for the electrons are the momentum in the $y$-direction $k=\frac{2\pi}{L}n$, where $n$ is an integer (this determines the position in the $x$-direction) and the spin $\sigma=\uparrow, \downarrow$. We take $k=0$ at the right edge; the left edge is then at $-k_F$, where $k_F=\frac{2\pi}{L}(N-1)$ is the Fermi momentum. The geometry is displayed in Fig.~\ref{fig_1}.

The Hamiltonian for our system is
{\setlength\arraycolsep{2pt}
\begin{eqnarray}
\label{H}
 \hat{H} & = & \frac{1}{2} \sum_{kpp' \sigma \sigma'} V(p,p')c_{k+p',\sigma}^{\dag}c_{k+p-p',\sigma'}^{\dag}c_{k+p,\sigma'}c_{k,\sigma} - \sum_{k \sigma,-k_F \le p \le 0} V(p-k,0) c_{k\sigma}^{\dag}c_{k\sigma} \nonumber \\
& & + \sum_{k \sigma} \delta V(k) c_{k\sigma}^{\dag}c_{k\sigma} + \sum_{k} g \mu_B B(c_{k\uparrow}^{\dag}c_{k\uparrow}-c_{k\downarrow}^{\dag}c_{k\downarrow}).
\end{eqnarray} }
The first term describes the Coulomb interaction between electrons in the gas, the second is the interaction of the electrons with an in-plane compensating positive background charge. $\delta V(k)$ is an additional potential (to be discussed below), and the last term is the Zeeman energy. (A constant term corresponding to the self-interaction of the background charge density has been dropped.) In the LLL, the cyclotron energy $\frac{1}{2}\hbar \omega_c$ is common for all electrons, and is ignored. The second and the third term together form the potential confining the electrons to the sample. The case $\delta V(k)=0$ is called the ideal edge. $c_{k\sigma}^{\dag}$, $c_{k\sigma}$ are the creation and annihilation operators for electrons with quantum numbers $k$, $\sigma$. $V(p,p')$ are the matrix elements for the unscreened Coulomb interaction $U(r)=e^2/r$. Using (\ref{psi}), we find
\begin{equation}
\label{me}
V(p,p') = \frac{e^2}{L} \sqrt{\frac{2}{\pi}} \exp^{-[\frac{p'^2}{2}+\frac{(p-p')^2}{2}]} \int_{-\infty}^{\infty} dy K_0(|p'y|) e^{-(y^2+2(p-p')y)/2}.
\end{equation}

For the ideal edge, the groundstate of the 2DEG is the compact ferromagnetic $\nu=1$ state 
\begin{equation}
\label{v1state}
|\nu=1 \rangle = \prod_{-k_F \le k \le 0} c_{k\uparrow}^{\dag}|0 \rangle,
\end{equation}
i.e. all the spin up states within $-k_F \le k \le 0$ are occupied.

We consider particle-hole (ph) excitations of (\ref{v1state}) within the LLL. They are characterized by two conserved quantum numbers: the excitation momentum $q$ and the spin $s=0,1$. The states are
\begin{eqnarray}
\label{exc}
|q,s=0 \rangle = \sum_{-q \le k \le 0} \phi_{k\uparrow}c_{k+q\uparrow}^{\dag}c_{k\uparrow} |\nu=1 \rangle, \\
|q,s=1 \rangle = \sum_{-k_F \le k \le 0} \phi_{k\downarrow}c_{k+q\downarrow}^{\dag}c_{k\uparrow} |\nu=1 \rangle.
\end{eqnarray}
The range of the momentum for the $s=0$ excitation is restricted by the Pauli principle. The wave functions $\phi_{k \sigma}$, and the corresponding eigenvalues (which determine the excitation energies) are calculated by numerical diagonalisation of the Hamiltonian (\ref{H}) in the ph-subspace $\{ c_{k\sigma}^{\dag}c_{p\uparrow}|\nu=1 \rangle \}$. For $s=0$, one finds the EMP mode, and for $s=1$, one finds bulk spin waves as well as spin waves bound to the edge (ESW).

\section{The edge potential}

The electrons in the gas are confined by the edge potential. The Coulomb repulsion acts to expand the gas, and it is clear that if the potential softens (compared to the ideal edge potential), the gas will eventually undergo a phase transition to a new groundstate. $\delta V(k)$ is used to vary the confining potential, with $-\hat{k}\frac{d}{dk}\delta V(k)$ as the corresponding force determining the softness.

\begin{figure}[!htb]
\begin{center}
\includegraphics[angle=270,width=0.49\textwidth]{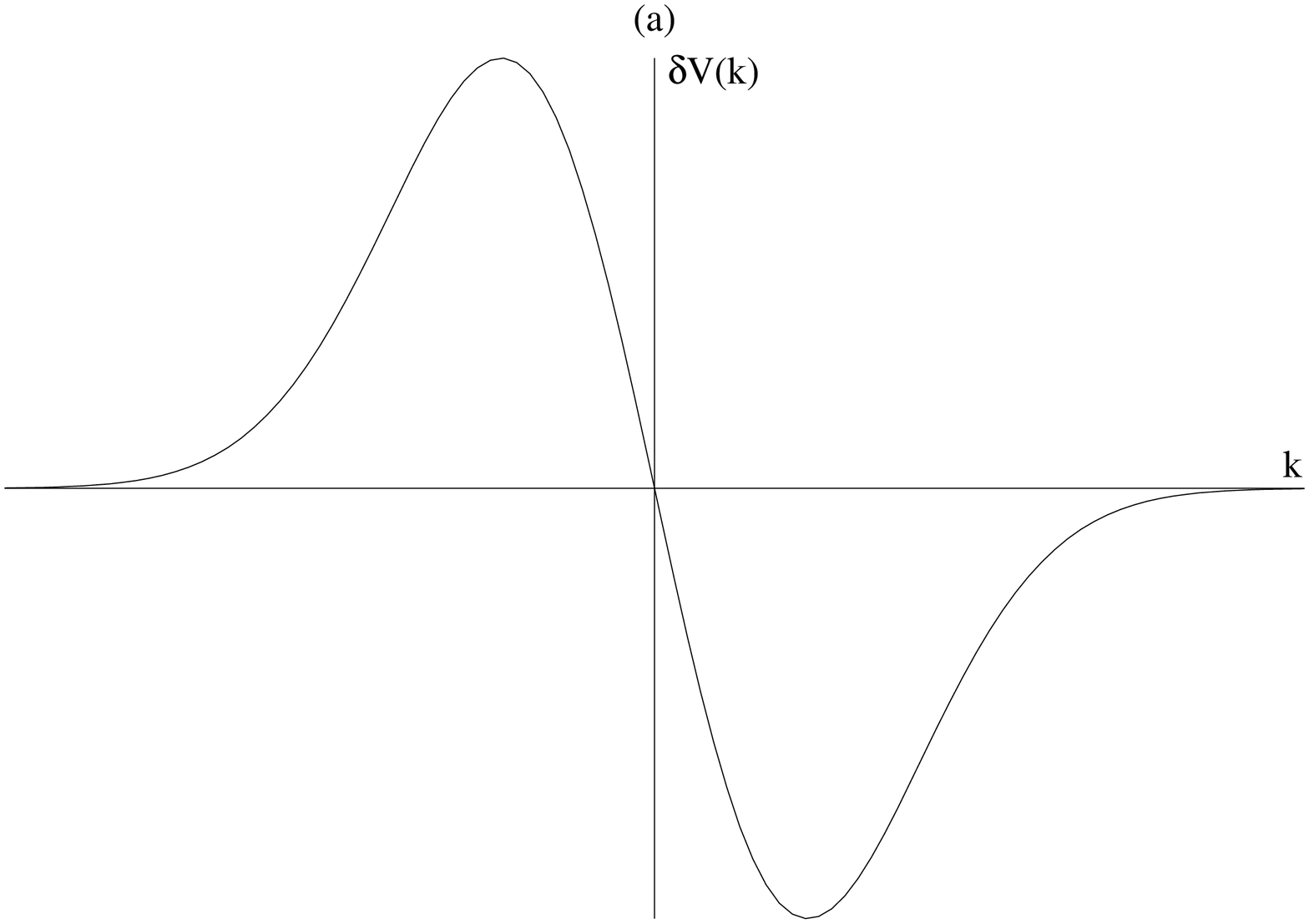}
\includegraphics[angle=270,width=0.49\textwidth]{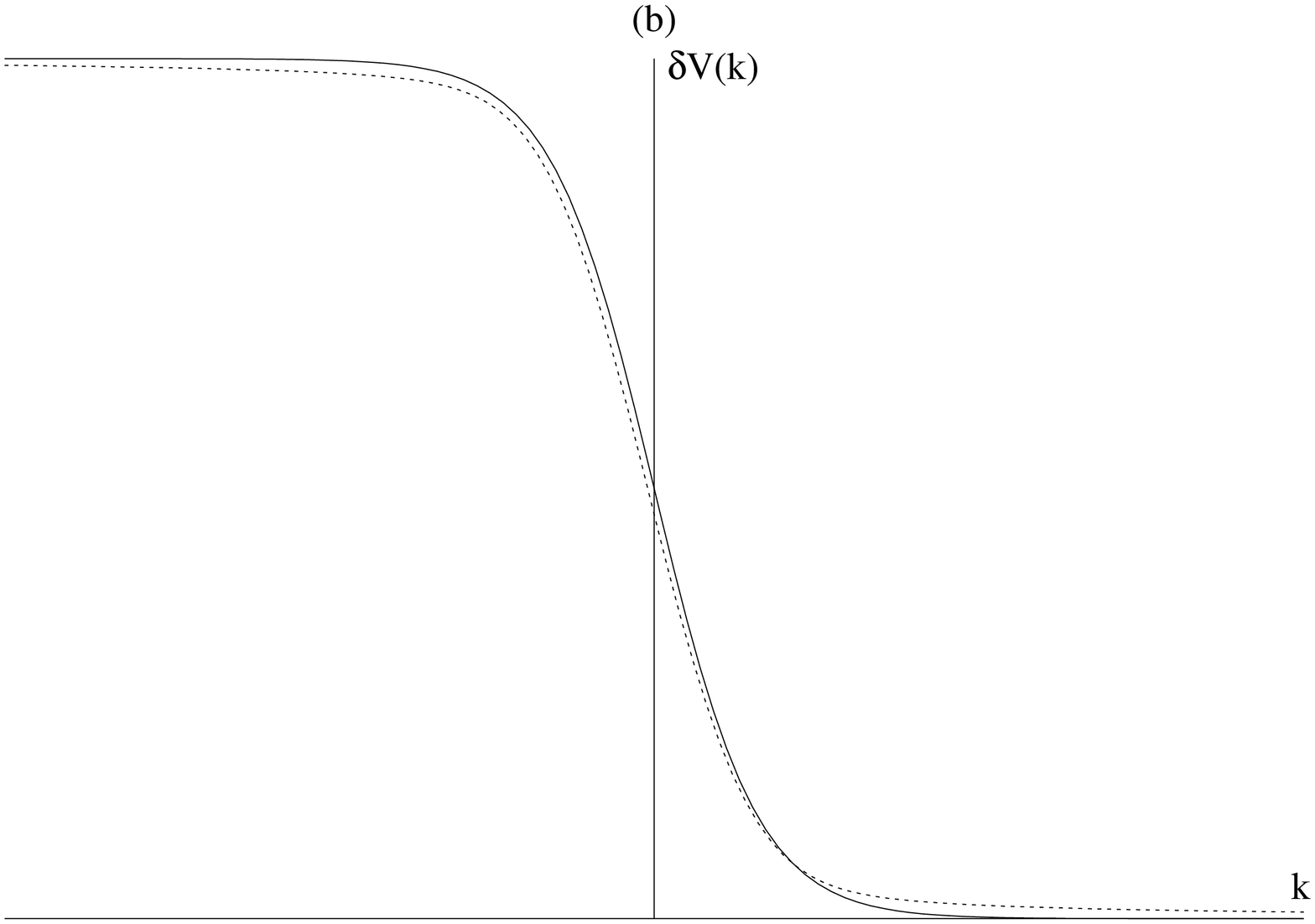}
\caption{\footnotesize{The three different potentials. (a) $\delta V_w(k)$, $w$ is approximately the width of the linear region around $k=0$. (b) $\delta V_{DCD}(k)$ is the dotted curve, $\delta V_{th}(k)$ is the solid one. The asymptotic value as $k \to -\infty$ is $d$.}}
\label{Pots}
\end{center}
\end{figure}

A standard model for the confinement is the $w$-model \cite{Chamon}. Here the confining potential is obtained from a compensating positive background charge density that decreases linearly from its bulk value to zero over a width $w$ centered at $k=0$. $w=0$ corresponds to the ideal edge. For large enough $w$, phase transitions are observed. In this model, $\delta V(k)$ becomes
\begin{equation}
\label{kl_pot}
\delta V_w(k) = \sum_p V(p-k,0) \rho_w(p),
\end{equation}
where
\begin{displaymath}
\rho_w(p) = \left\{ \begin{array}{ll}
p/w + 1/2 & \textrm{$-w/2 \leq p \leq 0$}\\
p/w - 1/2 & \textrm{$0 < p \leq w/2$}\\
0 & \textrm{$|p| > w/2$}.\\
\end{array} \right.
\end{displaymath}
For a typical $\delta V_w(k)$, see Fig.~\ref{Pots}a. The $w$-model turns out to be a non-generic choice for the confining potential, due to the property $\delta V_w(-\infty)=\delta V_w(0)$ \cite{KL5}.

A more realistic model for the confinement may be obtained by displacing the $w=0$ charge distribution a distance $d$ from the plane of the 2DEG - the displaced charge distribution (DCD) model. This models the displacement of the doped semiconductor layer present in real samples. In this case, one obtains the potential \cite{AE}
\begin{equation}
\label{ae_pot}
\delta V_{DCD}(k) = - \frac{1}{2\pi \sqrt{\pi}} \int_{-\infty}^{\infty} dx \ e^{-(x-k)^2} \Big[ 2d \arctan \frac{x}{d} + x \ln \frac{x^2+d^2}{x^2} - \pi d \Big].
\end{equation}
(To derive this expression we calculated the total potential from two semi-infinite charge planes, at a point in the 2DEG plane.) $\delta V_{DCD}(k)$ converges to its asymptotic value in the bulk, $d$, very slowly. This makes the numerical results dependent on the left edge of the sample. To overcome this finite size effect, we replace the DCD potential by 
\begin{equation}
\label{tanh_pot}
\delta V_{th}(k) = \frac{d}{2} \Big( 1 - \tanh \left( k \right) \Big).
\end{equation}
$\delta V_{th}(k)$ is very similar to $\delta V_{DCD}(k)$, as is seen in Fig.~\ref{Pots}b. By construction, it has the same asymptotic values, $\delta V_{th}(\infty) = 0$ and $\delta V_{th}(-\infty) = d$, as the DCD potential, but the convergence is exponential. The parameter $d$ is used to vary the potential softness; increasing $d$ means increasing the softness since the force $-\hat{k}\frac{d}{dk}\delta V(k) =\frac{1}{2} d\cosh^{-2}(k) \hat{k}$ is linear in $d$ and acts in the $+\hat{k}$ direction.

\section{Numerical analysis}

As seen from the Hamiltonian, the energies characterizing the system are the Coulomb energy $e^2/\epsilon l$ (here we allow for arbitrary $\epsilon$, this can be included in previous expressions by replacing $e^2$ with $e^2/\epsilon$), the Zeeman energy $g \mu_B B$ and the energy from the confining potential. We define $\tilde{g} = \frac{g\mu_B B}{e^2 / \epsilon l}$ as the dimensionless parameter determining the bulk physics. Note that the ESW mode energy is linear in $\tilde{g}$, whereas the EMP is gapless.

For a semi-infinite system ($k_F \to \infty$), the dispersion relation $\epsilon(q)$ of the ESW mode is known for $q \ll 1$ \cite{KL5},
\begin{equation}
\label{ESWthdisp2}
\epsilon (q) =  \frac{1}{4} \sqrt{\frac{\pi}{2}}q^2 - \sqrt{\frac{32}{\pi}} \Big( \delta V(-\infty) - \delta V(0) \Big)^2 q^2 + O(q^4),
\end{equation}
here, $\tilde{g}=0$. We now see explicitly what is special about the $w$-model; the $q^2$-term is unaffected by $\delta V_w(k)$. With $\delta V_{th}(k)$ as the potential in (\ref{ESWthdisp2}), we get $\delta V_{th}(-\infty) - \delta V_{th}(0) = d/2$ and 
\begin{equation}
\label{ESWdispf}
\epsilon_d (q) = \left( \frac{1}{4} \sqrt{\frac{\pi}{2}} - \sqrt{\frac{2}{\pi}}d^2 \right)q^2 + O(q^4)
\end{equation}
for the energy. Here we have added an index $d$ to emphasize the dependence on the confining potential. Eq. (\ref{ESWdispf}) can be used to test our calculations. For $q\ll 1$ at a given $d$, our numerical values should converge to the energy given by $\epsilon_d(q)$ when the system size increases.

We want to calculate energies for a half-plane, i.e. in the limit $N/L \to \infty$ (the width of the system is $k_F \propto \frac{N}{L}$) and $L \to \infty$. Upper limits for the parameters are set by the simulation time, which grows dramatically with $N$. In our simulations we varied $N$ between 200 and 800 and $L$ between 50 and 400, with $1.5 \leq N/L \leq 6$. To find the asymptotical values, we scale in $L/N$ and $1/L$ and extrapolate to zero. (When scaling in $L/N$ we keep $1/L$ constant and vice versa.)

The energies of the ESW and EMP modes only change appreciable with $L/N$ for small $q$. (This is true for all $d$.) At $q \approx 0.6$ the energies differ by $\sim 10^{-6}$ for $L/N$ between $0.17$ and $0.5$ (at constant $L$). For larger $L/N$, the difference increases. It also grows with decreasing $q$. At $q \approx 0.1$ for instance, the difference is $\sim 10^{-3}$ for $L/N$ between $0.17$ and $0.5$. For $q<0.4$, we use scaling in both $L/N$ and $1/L$. For $0.4<q<0.6$, we use $1/L$ scaling with the restriction $L/N \leq 0.5$, whereas for $q>0.6$, system sizes down to $L/N=2/3$ were used.

\begin{figure}[!htb]
\begin{center}
\includegraphics[angle=270,width=0.49\textwidth]{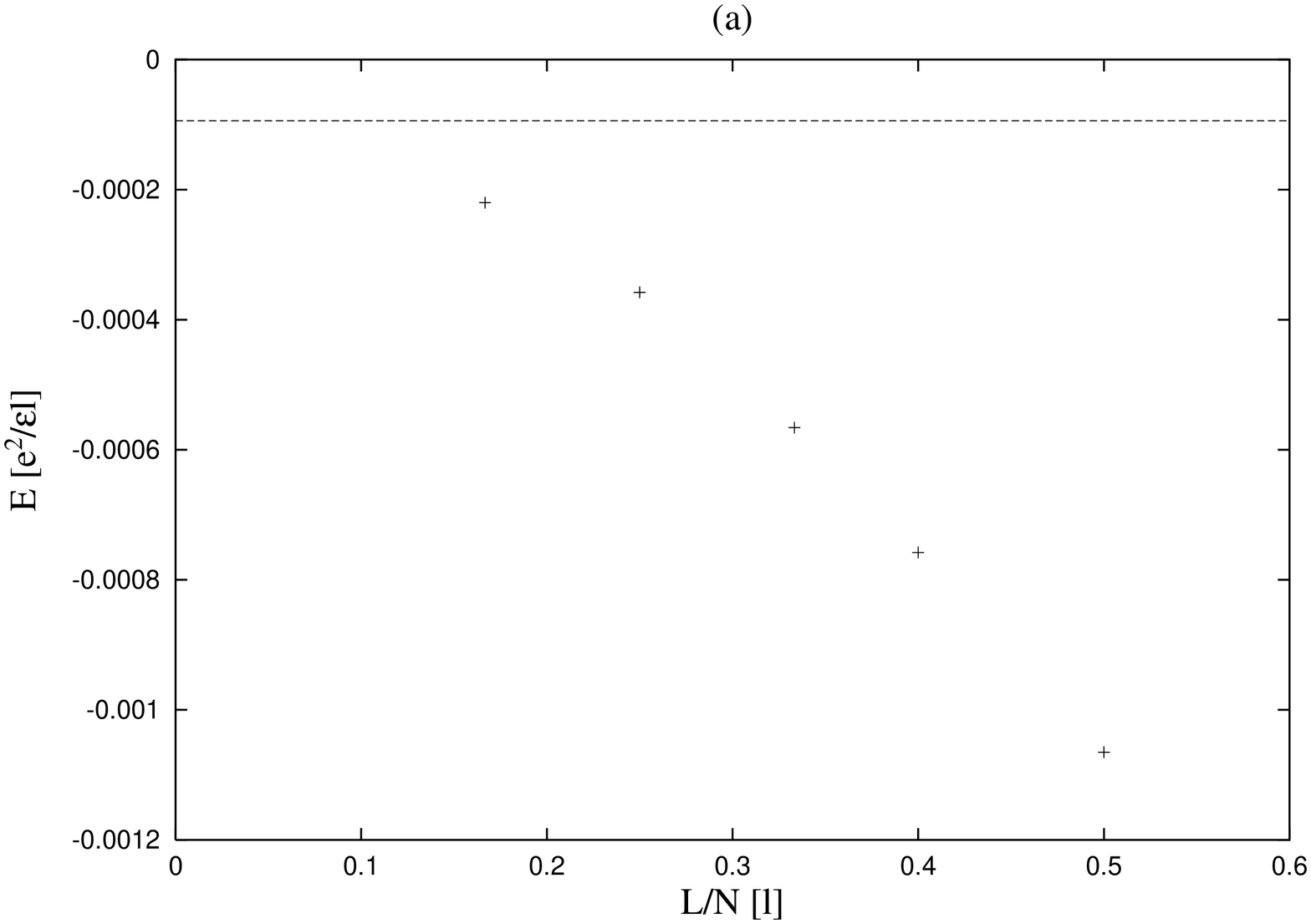}
\includegraphics[angle=270,width=0.49\textwidth]{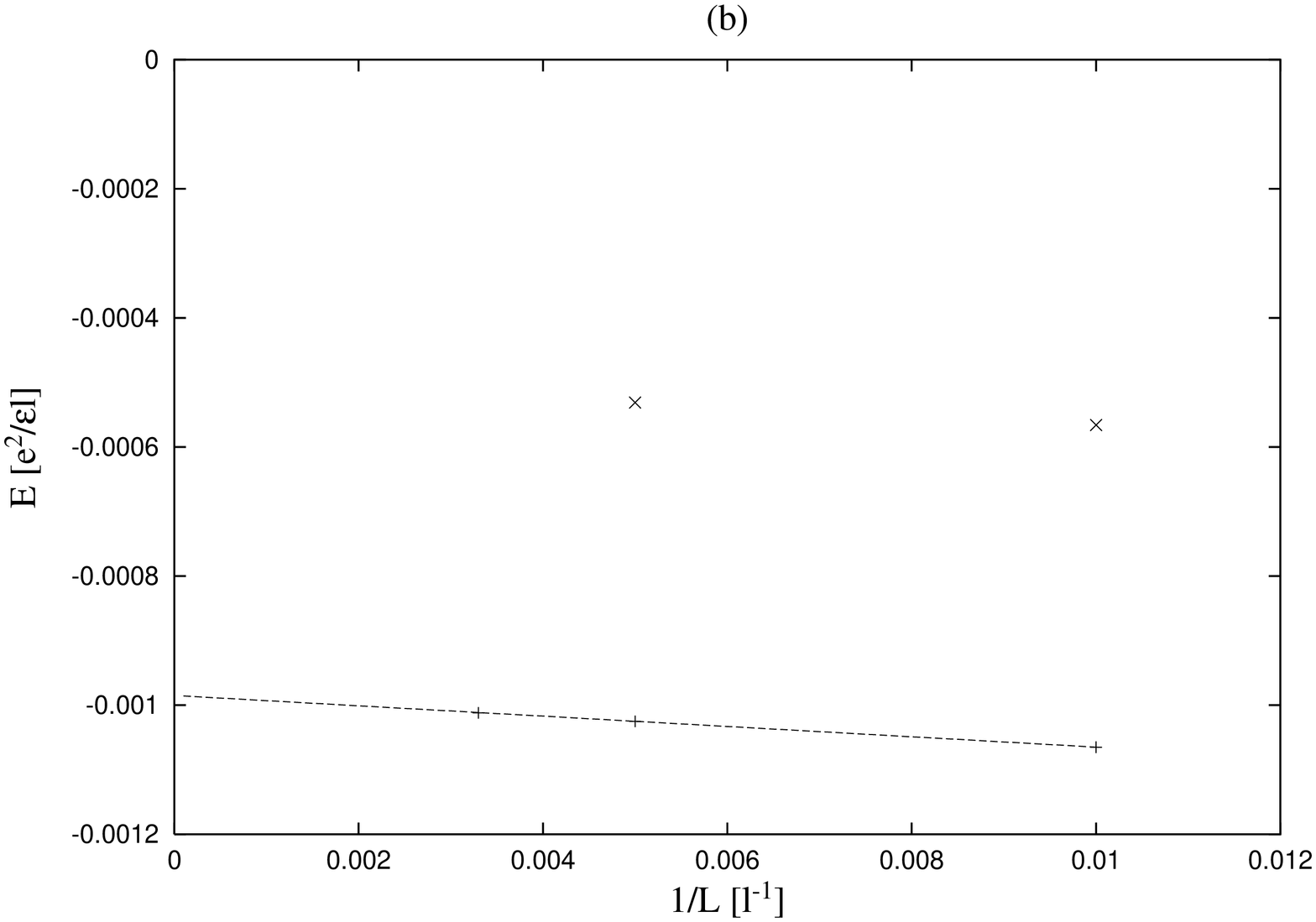}
\caption{\footnotesize{Scaling diagrams for the ESW at $d=0.65$ and $q=0.0628$. (a) The energy as a function of $L/N$, with $L=100$. The data points are from simulations, and the dashed curve is the $O(q^2)$-value predicted by (\ref{ESWdispf}). (b) The energy as a function of $1/L$. The upper data has $L/N=0.33$ and the lower $L/N=0.50$. The line is a least-squares fit.}}
\label{ESW_test}
\end{center}
\end{figure}

First we test our results at $q \ll 1$. Fig.~\ref{ESW_test} shows the ESW energy as a function of $L/N$ and $1/L$ at $d=0.65$ and $q=0.0628$, compared to the $q^2$-term of (\ref{ESWdispf}).
In Fig.~\ref{ESW_test}a the energy is seen to approach the predicted value nicely. (We do not expect an exact correspondece since we have a finite $q$.) In Fig.~\ref{ESW_test}b we see that the effect of increasing $L$ is marginal, certainly when compared to the energy variation in Fig.~\ref{ESW_test}a. This analysis can be repeated for different values of $d$, and the agreement with the model in (\ref{ESWdispf}) is good at least for $0.05 \leq d \leq 0.80$. We conclude that our numerical results are trustworthy.

\begin{figure}[!htb]
\begin{center}
\includegraphics[angle=270,width=0.49\textwidth]{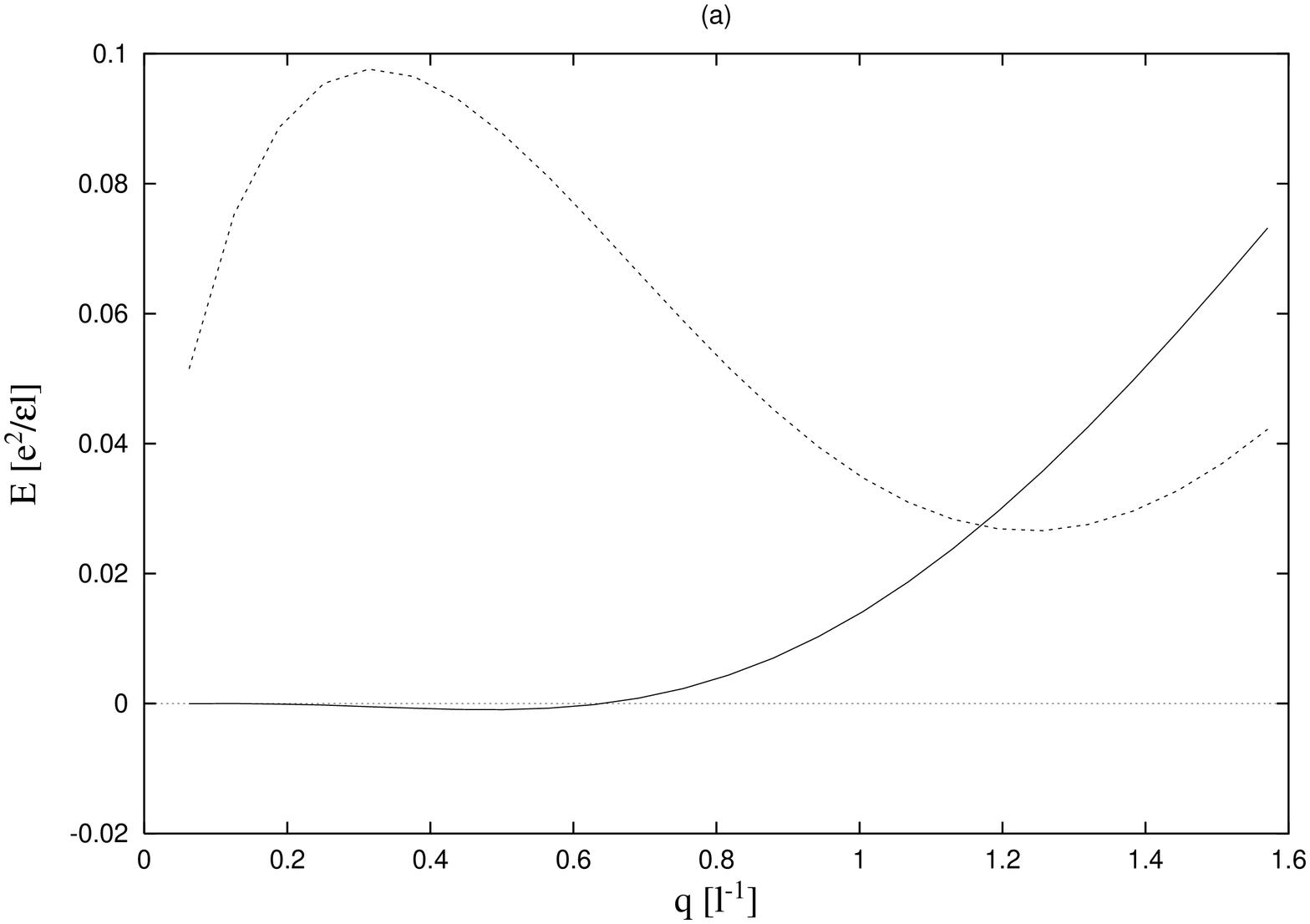}
\includegraphics[angle=270,width=0.49\textwidth]{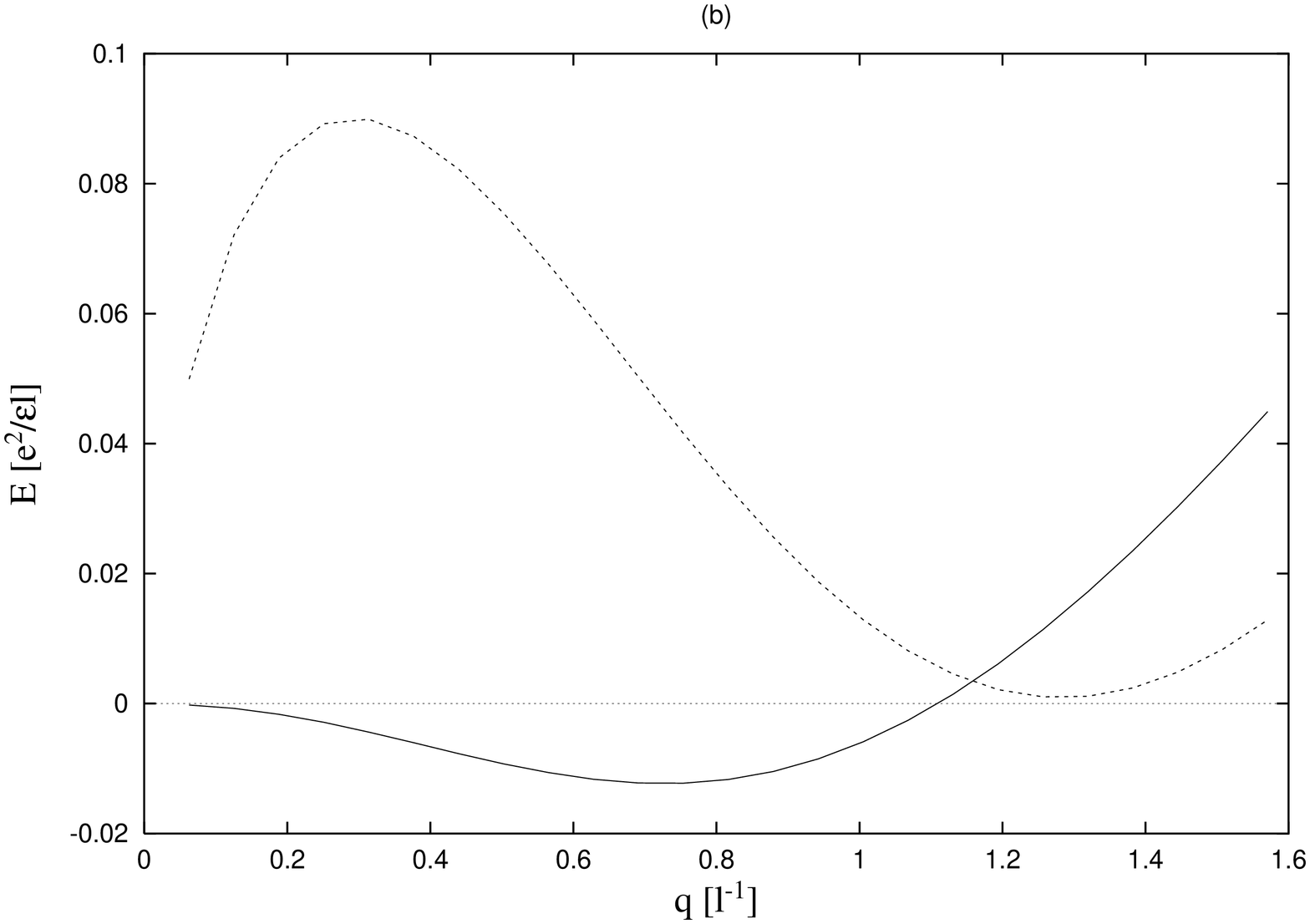}
\caption{\footnotesize{Dispersion relations for the two excitations, at $\tilde{g}=0$, $N=600$ and $L=100$. The solid curve is the ESW mode and the dotted the EMP. In (a) $d=0.600$, and in (b) $d=0.650$.}}
\label{Disp_rel}
\end{center}
\end{figure}

To find the critical points where the modes go soft, we first isolate the approximate $d$ and $q$ values where this happens. Fig.~\ref{Disp_rel} shows typical results for the dispersion relations for both types of excitations for two different values of $d$, with $\tilde{g}=0$, $N=600$ and $L=100$.

\begin{figure}[!htb]
\begin{center}
\includegraphics[angle=270,width=0.60\textwidth]{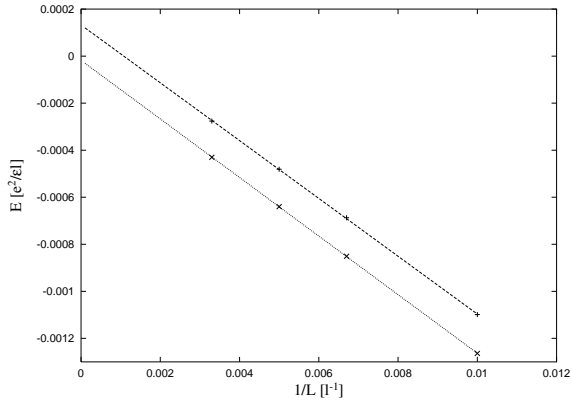}
\caption{\footnotesize{Minimum energy for the ESW mode as a function of $1/L$. The upper data points are at $d=0.601$, the lower at $d=0.602$. The lines are least-squares fits.}}
\label{d_esw}
\end{center}
\end{figure}

In Fig.~\ref{Disp_rel}a we are close to the point where the ESW mode goes soft. We determine the $d$ value more exactly and in Fig.~\ref{d_esw} we see the minimum energy as a function of $1/L$ for two different values of $d$, with the minimum occuring at $q\approx 0.48$ (weakly depending on $L$ due to the quantisation of $q$).
We see that the energy varies quite strongly with $L$, however, a linear least-squares fit matches the data very well. Using this line to extrapolate we get $d_{ESW}=0.602$ for the critical value. Note that the analytic result, eq. (\ref{ESWdispf}), predicts a phase transition at $q=0$ for $d \approx 0.627$ (if only the $q^2$-term is considered), whereas we find the transition for finite $q$, where $\epsilon_d(q)$ does not apply, and a \textit{smaller} $d$. We can thus safely conclude that we have found the correct minimum.

\begin{figure}[!htb]
\begin{center}
\includegraphics[angle=270,width=0.60\textwidth]{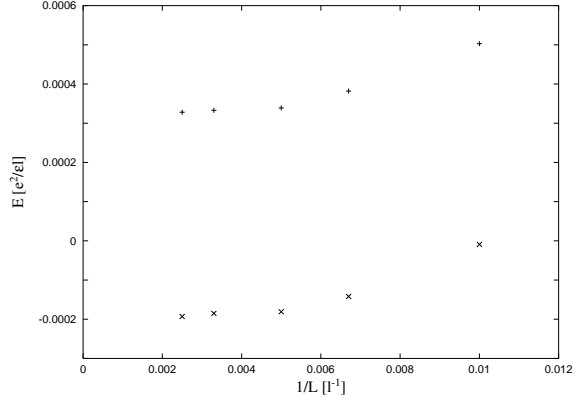}
\caption{\footnotesize{Minimum energy for the EMP mode as a function of $1/L$. The upper data have $d=0.651$, the lower $d=0.652$.}}
\label{d_emp}
\end{center}
\end{figure}

In Fig.~\ref{Disp_rel}b, $d \approx d_{EMP}$, where the EMP mode goes soft. We examine the energy at the minimum ($q \approx 1.28$) for the EMP mode as a function of $1/L$ for two different values of $d$ in Fig.~\ref{d_emp}.
We clearly see that $d_{EMP}=0.652$ is the best approximation.

\begin{figure}[!htb]
\begin{center}
\includegraphics[angle=270,width=0.60\textwidth]{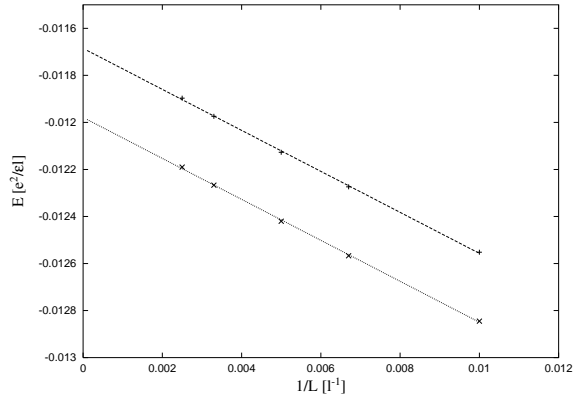}
\caption{\footnotesize{Minimum energy for the ESW mode as a function of $1/L$. The upper data points are at $d=0.651$, the lower at $d=0.652$. The lines are least-squares fits.}}
\label{gc}
\end{center}
\end{figure}

We also see from Fig.~\ref{Disp_rel}b that the ESW mode has continued to drop below zero as we increased $d$. In order to have a CDW phase, we need the ESW mode energies to be positive. Since the ESW mode is linear in $\tilde{g}$, the energies will become positive if $\tilde{g} \geq \tilde{g}_c$, where $\tilde{g}_c$ is the absolute value of the minimal ESW energy, at $d=d_{EMP}$. Fig.~\ref{gc} shows the minimum ESW energy (at $q \approx 0.75$) as a function of $1/L$, for $d=0.651$ and $0.652$.
The $L$ dependence is fairly strong, but again the data is practically linear in $1/L$. We already know that $d_{EMP}$ is closer to $0.652$ than $0.651$, and, as we can see, $\tilde{g}_c$ only changes by $\sim 10^{-4}$ in that interval. We conclude that $\tilde{g}_c=0.012$.

\section{Summary}

Along the curve from $(0,d_{ESW})$ to $(\tilde{g}_c,d_{EMP})$ in $\tilde{g}d$-space there is a phase transition into a TEX edge. Similarly, along the line $d=d_{EMP}$ and $\tilde{g} \geq \tilde{g}_c$ there is a phase transition into a CDW edge. The critical values are $d_{ESW}=0.602$, $d_{EMP}=0.652$ and $\tilde{g}_c =0.012$.

\begin{figure}[!htb]
\begin{center}
\includegraphics[angle=270,width=0.7\textwidth]{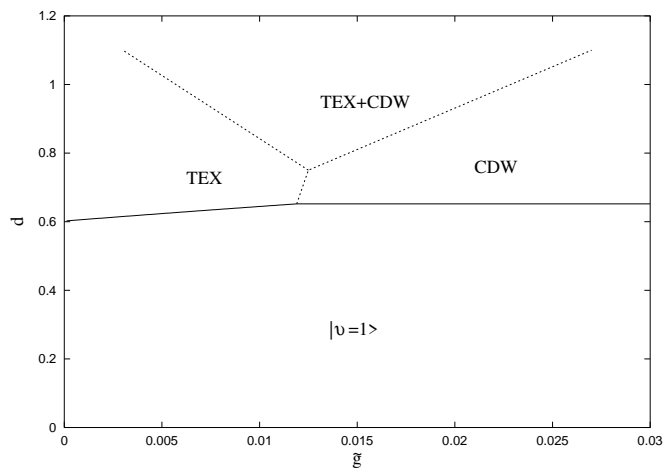}
\caption{\footnotesize{Phase diagram in $\tilde{g}d$-space with the different regions corresponding to different groundstates. Solid lines are numerically determined, dotted ones are only topologically known.}}
\label{phd}
\end{center}
\end{figure}

In our calculations we assume that the groundstate of the edge is the compact ferromagnetic $|\nu=1 \rangle$ state, and, therefore, we cannot determine anything beyond the TEX and the CDW phase transitions. At larger $d$, however, a phase that is a combination of the two is expected \cite{Franco2}. Fig.~\ref{phd} shows the phase diagram.

For the $w$-model, using the explained numerical techniques, we obtain the asymptotical value $\tilde{g}_{c,w}=0.008$ for $w=7.03$. Thus the $w$-model underestimates the region where the transition is into the TEX edge.

\begin{center}
\Large{\bf Acknowledgments }\\
\end{center}
We wish to thank K. Lejnell for providing Hartree-Fock results and D. Lillieh\"o\"ok, S. Isakov and J. Br\"annlund for their computer skills. Anders Karlhede was supported by the Swedish Science Research Council.\\

\end{document}